\documentclass[12pt]{article}
\usepackage{amsmath,amsfonts,amssymb}
\usepackage{graphicx}
\usepackage{psfrag}
\usepackage{enumerate}

\numberwithin{equation}{section}

\begin{document}
\newcommand{\todo}[1]{{\em \small {#1}}\marginpar{$\Longleftarrow$}}
\newcommand{\labell}[1]{\label{#1}\qquad_{#1}} 

\rightline{DCPT-06/15}
\vskip 1cm

\begin{center}
{\Large \bf Boundary Conditions and Dualities: \\ Vector Fields in
AdS/CFT}
\end{center}
\vskip 1cm

\renewcommand{\thefootnote}{\fnsymbol{footnote}}
\centerline{\bf Donald Marolf\footnote{marolf@physics.ucsb.edu} and Simon
F. Ross\footnote{S.F.Ross@durham.ac.uk}}
\vskip .5cm
\centerline{ \it Department of Physics, University of California}
\centerline{ \it Santa Barbara, CA 93106, U.S.A.}
\vskip .5cm
\centerline{ \it Centre for Particle Theory, Department of
Mathematical Sciences}
\centerline{\it University of Durham, South Road, Durham DH1 3LE, U.K.}

\setcounter{footnote}{0}
\renewcommand{\thefootnote}{\arabic{footnote}}

\vskip 1cm
\begin{abstract}
  In AdS, scalar fields with masses slightly above the
  Breitenlohner-Freedman bound admit a variety of possible boundary
  conditions which are reflected in the Lagrangian of the dual field
  theory.  Generic small changes in the AdS boundary conditions correspond to
  deformations of the dual field theory by multi-trace operators.
  Here we extend this discussion to the case of vector gauge fields in
  the bulk spacetime using the results of Ishibashi and Wald [hep-th/0402184].
  As in the context of scalar fields, general boundary conditions
  for vector fields involve multi-trace deformations which lead to
renormalization-group flows.  Such flows originate in ultra-violet
CFTs which give new gauge/gravity dualities. At least for
AdS$_4$/CFT$_3$, the dual of the bulk photon appears to be a
propagating gauge field instead of the usual R-charge current.
Applying similar reasoning to tensor fields suggests the existence
of a duality between string theory on AdS${}_4$ and a quantum
gravity theory in three dimensions.
\end{abstract}

\section{Introduction}

In the AdS/CFT correspondence, boundary conditions for bulk fields are
related to the specification of the dual
CFT~\cite{Maldacena:1997re,Witten:1998qj,Gubser:1998bc,Aharony:1999ti}.
In particular, small changes in the bulk boundary conditions
correspond to deformations of the dual CFT Lagrangian. Bulk scalar
fields in AdS$_{d+1}$ with mass in the range $-d^2/4 \leq m^2 < -d^2/4
+1$ provide a particularly interesting example of this correspondence.
As indicated by the work of Breitenlohner and
Freedman~\cite{Breitenlohner:1982bm,Breitenlohner:1982jf}, such scalar
fields admit a variety of possible boundary conditions.  In
particular, one may fix either the faster or slower falloff part of
the scalar field at infinity.

The two resulting bulk theories correspond to two different dual
CFTs, in which the field $\phi$ is dual to operators of dimensions
$\Delta_-$ and $\Delta_+ = d-\Delta_-$ respectively, where $d/2
\geq \Delta_- > d/2-1$. In~\cite{Witten:2001ua,Berkooz:2002ug}, it
was observed that a general linear boundary condition, relating
the faster falloff part to the slower, corresponds to a
double-trace deformation, adding a term $f \mathcal{O}^2$ to the
Lagrangian of the CFT. Starting from the $\Delta_-$ CFT, this is
is a relevant deformation, which will produce an
renormalization-group flow which is expected to end at the
$\Delta_+$ CFT in the IR; evidence for this picture has been
obtained in~\cite{Gubser:2002zh,Gubser:2002vv,Hartman:2006dy}.
Since a double-trace operator corresponds to a multiparticle
state, the double-trace deformations in the CFT have also been
related to worldsheet non-locality in the bulk string
theory~\cite{Aharony:2001pa,Aharony:2001dp}.

In the present work, we conduct a similar analysis for vector
fields.  The possibility of general boundary conditions for vector
gauge fields was first raised in \cite{Breitenlohner:1982jf} for
$d=3$. In a recent thorough analysis by Ishibashi and
Wald~\cite{Ishibashi:2004wx}, it was shown that for
electromagnetic and gravitational perturbations in AdS spacetime,
both the slow- and fast- falloff pieces of certain parts of the
field are normalizable for $d=3,4,5$; i.e., for bulk spacetime
dimensions 4, 5 and 6. As a result, these fields admit general
classes of boundary conditions. We investigate the dual CFT
description of such general theories, focusing on the
electromagnetic perturbations for simplicity. As in the scalar
case, we will find different CFTs corresponding to fixing the
faster and slower falloff pieces of the bulk field. Furthermore, a
general local linear boundary condition corresponds to a
deformation of the former theory by a relevant operator,
generating a renormalization-group flow which should lead to the
latter.

However, a number of interesting new features arise in the vector
case. Some of these are associated with gauge invariance.  In the
slow falloff CFT, the operator dual to the bulk photon is a CFT
gauge field instead of the more familiar $R$-symmetry current.  As
a result, a general boundary condition is dual to a field theory
for which the gauge-invariant action is non-local, though it
becomes local in the gauge picked out by the boundary condition.
Other features have to do with the possibility of deforming only
certain pieces of the gauge field, breaking Lorentz invariance as
a result.

After posting the first version of this paper on the hep-th arxiv,
we became aware of a body of literature with results overlapping
those presented here for the case $d=3$. In particular, the fact
that `conjugate' boundary conditions in AdS$_4$ are dual to a
CFT$_3$ with a dynamical gauge theory was described in
\cite{Witten:2003ya} and discussed further in
\cite{Leigh:2003gk,Leigh:2003ez,Petkou:2004nu,Yee:2004ju}. Certain
aspects of the general multi-trace deformations and
renormalization group flows were discussed in
\cite{Leigh:2003gk,Leigh:2003ez,Petkou:2004nu}, and these
references also study higher spins for $d=3$.  For spin 1, the
conjugate CFT is related to the quantum Hall effect
\cite{Burgess:2000kj}. For higher spins, there is a relation to
higher spin theories in AdS$_4$; see
\cite{Vasiliev:2004cp,Bouatta:2004kk} for recent reviews.  This
earlier work focuses largely on the CFT point of view; our work
provides a bulk counterpart, considers certain details required to
yield a fully local theory, and addresses extensions to $d=4,5$, and the allowed boundary condition for $d=2$. 

We begin by carefully reviewing the analysis of the scalar case in
section \ref{scalar}.  We then address boundary conditions for
vector gauge fields in section~\ref{vecbcsec}, drawing heavily on
the results of~\cite{Ishibashi:2004wx}. In section~\ref{cft}, we
develop our proposal for the dual CFT description.
Some final remarks concerning both vector fields and
extrapolations to tensor fields are contained in
section~\ref{disc}.

\section{Scalar fields: general linear boundary conditions and
  double-trace deformations}
\label{scalar}

This section reviews the relation between boundary conditions for
scalar fields and the associated deformations of the dual field
theory.  This correspondence was conjectured in
\cite{Witten:2001ua,Berkooz:2002ug}, derived in \cite{Sever:2002fk},
and studied further in, e.g.
\cite{Minces:2001zy,Muck:2002gm,Minces:2002wp,Minces:2004zr}.  Our
treatment below is essentially a Lorentzian version of
\cite{Sever:2002fk}, extended in section \ref{satBF} to the case of
scalars with logarithmic behavior near the boundary of AdS.  For
simplicity, we use the familiar toy model of AdS/CFT in which the bulk
theory is replaced by a real scalar test field $\phi$ in
AdS${}_{d+1}$.
\subsection{Scalars with $m^2 > m^2_{BF}$}

As stated above, we consider a real scalar field which propagates on a
fixed spacetime.  We take this spacetime to be AdS${}_{d+1}$, with AdS
length scale $\ell = 1$.  It is convenient to use coordinates such
that the AdS${}_{d+1}$ metric is
\begin{equation}
ds^2 = g_{ab} dy^a dy^b =  - (1+ r^2) dt^2 +  \frac{dr^2}{1 + r^2} +
r^2 d\Omega_{d-1}^2,
\end{equation}
where $d\Omega_{d-1}^2$ is the round metric on the unit $S^{d-1}$.

Since we are interested in boundary conditions, we first describe
the asymptotic behavior of the field. Suppose that our scalar is
associated with a potential $V(\phi)$ with squared mass $m^2 =
\frac{1}{2} V''(0)$. We restrict attention here to the case where
the mass is close to, but slightly above, the
Breitenlohner-Freedman
bound~\cite{Breitenlohner:1982bm,Breitenlohner:1982jf}:
\begin{equation}
\label{mrange}
-\frac{d^2}{4} +1 >
m^2 >  - \frac{d^2}{4}.
\end{equation}
For such values of $m$, one finds that all solutions to the equations
of motion take the asymptotic form
\begin{equation}
\label{asympt}
\phi \rightarrow \frac{\alpha(x)}{r^{\lambda_-}} +
\frac{\beta(x)}{r^{\lambda_+}},
\end{equation}
where $x$ are coordinates on null infinity ($\partial {\cal M}$, also known as the conformal boundary) and where
\begin{equation}
\lambda_\pm = \frac{d}{2} \pm \frac{1}{2} \sqrt{d^2 + 4 m^2}.
\end{equation}
Note that (\ref{mrange}) implies
\begin{equation}
\label{lrange}
2 > \lambda_+ - \lambda_- > 0.
\end{equation}
The case $m^2 = -d^2/4$ involves various logarithmic terms and will be
treated separately in section \ref{satBF} below.

The boundary condition should be chosen to yield a well-defined
phase space.  This occurs when the symplectic structure is finite
and the symplectic flux\footnote{The symplectic flux for a scalar
field is proportional to the Klein-Gordon flux.  See e.g.
\cite{WaldThermo,Lee:1990nz}, for general comments on symplectic
structures and their role in quantization.}  through infinity
vanishes, so that the symplectic structure is conserved.

The mass range (\ref{mrange}) is precisely the range for which all
solutions (\ref{asympt}) are normalizable with respect to the
symplectic structure (see e.g. \cite{Klebanov:1999tb}).  Thus, the
only constraint is the requirement that the flux through infinity
vanish.  For two vectors $\delta_1\phi, \delta_2\phi$ tangent to
the space of solutions, the symplectic flux through a region $R$
of null infinity is
\begin{equation}
\label{sflux} \omega_R(\delta_1 \phi, \delta_2 \phi) = (\lambda_+
- \lambda_-) \int_R \sqrt{\Omega} ( \delta_1 \alpha \delta_2 \beta
- \delta_1 \beta \delta_2 \alpha ).
\end{equation}
If our boundary condition is to force (\ref{sflux}) to vanish for all regions $R$, then $\alpha$ must be an ultra-local function of $\beta$; i.e., $\alpha(x)$ can depend only on $\beta(x)$ at a point, and cannot depend on derivatives of $\beta$:
\begin{equation}
\label{sgen}
\alpha (x) = J_\alpha(x, \beta) \ \ \ {\rm or} \ \ \ \beta (x) = J_\beta(x, \alpha)  .
\end{equation}
Note that in each case, vanishing of (\ref{sflux}) implies the existence of a potential $W_\alpha(\beta)$, $W_\beta(\alpha)$ such that
\begin{equation}
\frac{1}{\sqrt{\Omega}} \frac{\delta W_\alpha }{\delta \beta (x)}=
\label{spot} (\lambda_+ - \lambda_-) J_\alpha (x,\beta)  \ \ \  \frac{1}{\sqrt{\Omega}} \frac{\delta W_\beta }{\delta \alpha (x)}=
- (\lambda_+ - \lambda_-) J_\beta (x,\alpha) ,
\end{equation}
where the normalization factor $(\lambda_+ - \lambda_-) $ on the
right-hand side was chosen for later convenience.     One may
further show that all such boundary conditions remain valid when
the scalar field is coupled to gravity; see \cite{Amsel:2006uf}
for a general analysis and
\cite{Henneaux:2002wm,Henneaux:2004zi,Hertog:2004dr,Hertog:2004ns,Hertog:2005hm,Henneaux:2006hk}
for direct calculations. We recall the implications of various
choices of such boundary conditions for AdS/CFT
below\footnote{While it would not correspond to our usual notion
of a local bulk theory, one could choose to require the integrated
flux (\ref{sflux}) to vanish only for a certain family of regions
$R$. For example, if vanishing flux is required only for regions
bounded by $t=constant$ surfaces then the boundary condition
$J_\alpha(x,\beta)$ can be taken to be non-local in space (but
still ultra-local in time), so long as $\frac{\delta
J_\alpha(x)}{\delta \beta(y)}$ is an appropriately self-adjoint
operator; i.e., so long as the potential $W_\alpha$ continues to
exist.  Such
settings may also be of interest for AdS/CFT. Further
generalizations should also be possible if one is willing to add
extra boundary degrees of freedom.}.

\subsubsection{Fixing $\alpha$}
\label{fixa}

Because AdS is not globally hyperbolic, we must impose a boundary condition on the scalar field.
Let us first suppose that one fixes the leading behavior by choosing some fixed function $J_\alpha$ on $\partial {\cal M}$ and imposing
\begin{equation}
\label{aisJ}
\alpha(x) = J_\alpha(x), \ \ \ {\rm for} \  x \in \partial {\cal M}.
\end{equation}
The coefficient $\beta(x)$ is then to be determined from the equations of motion and the initial conditions which, for the moment, we take to be given by specifying fixed values of $\phi$ on $\Sigma_\pm$:
\begin{equation}
\label{pf}
\phi(x) = \phi_\pm(x) , \ \ \ {\rm for} \  x \in \Sigma_\pm.
\end{equation}

A valid action must be stationary on solutions.  In particular, we wish the action to be stationary under all variations which preserve the boundary conditions (\ref{aisJ}) and (\ref{pf}).
To this end, consider the action
\begin{equation}
\label{KWA}
S_{\alpha = const} = - \int_{\cal M} \left( \frac{1}{2}\partial
\phi^2 + V(\phi) \right) \sqrt{-g} - \frac{1}{2} \lambda_- \int_{\partial \cal M} \sqrt{-h} \phi^2,
\end{equation}
where ${\cal M}$ denotes a region of $AdS_{d+1}$ bounded to the past
and future by Cauchy surfaces $\Sigma_-, \Sigma_+$, though we abuse notation by continuing to use $\partial {\cal M}$ to denote only the boundary at null infinity. As noted in
\cite{Minces:2004zr}, the action (\ref{KWA}) is equivalent to the ``improved
action'' advocated by Klebanov and Witten (see equation (2.14) of
\cite{Klebanov:1999tb}) for configurations satisfying (\ref{asympt}).
In (\ref{KWA}), $h$ denotes the determinant of the (divergent) induced
metric on null infinity.

We now compute variations:
\begin{equation}
\label{varyex} \delta S_{\alpha = const}=  \int_{\cal M} \sqrt{-g} \left( \nabla^2
\phi - V'(\phi)\right)  \delta \phi  - \int_{\partial \cal M}  \sqrt{-h} (n^a
\partial_a \phi) \delta \phi  - \lambda_- \int_{\partial  \cal M} \sqrt{-h} \phi \delta \phi,
\end{equation}
where $n$ is the outward pointing unit normal to $\partial {\cal M}$ (i.e., with $n^a n^b g_{ab} = \pm1$) and we have used (\ref{pf}) to show that the boundary terms at $\Sigma_\pm$ vanish.  We have
\begin{eqnarray}
 \int_{\partial \cal M}  \sqrt{-h} (n^a
\partial_a \phi) \delta \phi   &=&  -\int_{\partial  \cal M}  \sqrt{\Omega}
(r^{\lambda_+ - \lambda_-} \lambda_- \alpha \delta \alpha + \lambda_- \alpha \delta \beta  + \lambda_+ \beta \delta \alpha), \cr
    \int_{\partial  \cal M}  \sqrt{-h} \phi \delta \phi &=&
 \int_{\partial  \cal M} \sqrt{\Omega} (r^{\lambda_+ - \lambda_-} \alpha \delta \alpha +  \alpha \delta \beta  + \beta \delta \alpha),
 \end{eqnarray}
where $\Omega$ is the determinant of the metric on the unit $S^{d-1}$ sphere, and  we have neglected terms which vanish in the $r \rightarrow \infty$ limit.  In particular, we have used the fact that $n^a \partial_a =   (\sqrt{r^2 +1})   \partial_r =   (r + O(r^{-1})) \partial_r$  and (\ref{lrange}).  As a result, one finds
\begin{equation}
\label{varyex2} \delta S_{\alpha = const} =  \int_{\partial  \cal M} \sqrt{-g} \left( \nabla^2
\phi - V'(\phi)\right)  \delta \phi   + (\lambda_+ - \lambda_-) \int_{\partial  \cal M}  \sqrt{\Omega}
  \beta \delta \alpha .
  \end{equation}
 Since (\ref{aisJ}) implies $\delta \alpha=0$, we see that (\ref{KWA}) indeed provides a valid variational principle for such boundary conditions.  A similar calculation shows that under the same boundary condition the action $S_{\alpha = const}$ is also finite when the equations of motion hold.

Now, the variation of a path integral with respect to some family of deformations may be taken to define an operator.  Furthermore, in the semi-classical limit, variations of the path integral are given by variations of the on-shell action.  Consider then the operator ${\cal O}_\alpha$ in the dual CFT whose matrix elements are given in this approximation by the variation of the bulk on-shell action with respect to $J_\alpha(x)$:

\begin{equation}
\label{scalarOa}
\langle {\cal O}_\alpha \rangle = \frac{1}{\sqrt{\Omega}} \frac{\delta S_{\alpha = const}}{\delta J_\alpha} = (\lambda_+ - \lambda_-) \beta.
\end{equation}
It is convenient to denote a generic matrix element by $\langle {\cal O}_\alpha \rangle$ and to leave implicit the specification of states between which the matrix element is computed.

The choice of states between which one computes the matrix element $\langle {\cal O}_\alpha \rangle$ determines the boundary conditions at $\Sigma_\pm$ and as well as additional boundary {\it terms} at $\Sigma_\pm$ which must be added to $S_{\alpha = const}$.  For simplicity, we have suppressed such details here. As discussed in  \cite{Marolf:2004fy}, the net result of adding the additional terms and altering the boundary conditions is that (\ref{varyex2}) is unchanged, though the solution on which (\ref{varyex2}) is evaluated depends on the choice of states.

\subsubsection{Fixing $\beta$}
\label{fixb}

For masses in the range (\ref{mrange}),
one may similarly consider a theory with boundary condition $\beta =J_\beta(x)$ \cite{Breitenlohner:1982bm,Breitenlohner:1982jf}.
An appropriately stationary action for such theories is given by
\begin{eqnarray}
S_{\beta = const} &=&
- \int_{\mathcal M} \left( \frac{1}{2}\partial
\phi^2 + V(\phi) \right) \sqrt{-g} + \int_{\partial  \cal M} \sqrt{-h} \phi n_I^a \partial_a \phi
+ \frac{1}{2} \lambda_-\int_{\partial  \cal M} \sqrt{-h} \phi^2 \cr  &=& S_{\alpha = const} - (\lambda_+ - \lambda_-) \int_{\partial \cal M} \sqrt{\Omega} \beta \alpha,
\end{eqnarray}
for which we have
\begin{equation}
\delta S_{\beta = const} =  \int_{\cal M} \sqrt{-g} \left( \nabla^2
\phi - V'(\phi)\right)  \delta \phi -   (\lambda_+ - \lambda_-) \int_{\partial \cal M}  \sqrt{\Omega} \alpha \delta \beta.
\end{equation}

In each such theory, there is an operator $O_\beta$ associated with deformations of $J_\beta$:
\begin{equation}
\label{scalarOb}
\langle {\cal O}_\beta \rangle = \frac{1}{\sqrt{\Omega}} \frac{\delta S_{\beta=const}}{\delta J_\beta} = -(\lambda_+ - \lambda_-) \alpha.
\end{equation}
As conjectured in \cite{Klebanov:1999tb} and discussed in detail in \cite{Gubser:2002vv}, the bulk theory with $\beta =0$ boundary conditions is dual to a CFT for which the generating functional for planar diagrams is related to that of the $\alpha =0$ theory.

\subsubsection{More general boundary conditions}
\label{sW}

Two particular classes of boundary conditions were considered above, defined by fixing either the value of $\alpha$ or $\beta$ on $\partial {\cal M}$.  We now wish to consider the more general boundary conditions (\ref{spot}), starting with the case defined by a potential $W_\alpha(\beta)$.
From (\ref{varyex2}) we see that with the boundary condition (\ref{sgen})
 the original action $S_{\alpha = const}$ (\ref{KWA}) is no longer stationary on solutions.  The full action must be of the form
\begin{equation}
\label{SSWa}
S_{W_\alpha} = S_{\alpha =  const} + B(\alpha).
\end{equation}
On-shell, and for fixed boundary conditions at $\Sigma_\pm$, we clearly have
\begin{equation}
\delta S_{W_\alpha} =  \int_{\partial \cal M}  \sqrt{\Omega}
\left[ (\lambda_+ - \lambda_-) \beta \delta \alpha  +
\frac{1}{\sqrt \Omega} \frac{\delta B}{\delta \alpha} \delta
\alpha \right],
\end{equation}
so we must choose $B$ to satisfy
\begin{equation}
\label{Beq}
\frac{\delta B}{\delta \alpha } =  - (\lambda_+ - \lambda_-) \beta \sqrt{\Omega}.
\end{equation}

Let us now ask about the field theory dual of the bulk theory defined by the general boundary condition (\ref{sgen}).  The action of this theory will differ from the action $S^{FT}_{\alpha = 0}$ of the $\alpha=0$ CFT by some term $\Delta S^{FT}$. One may calculate how such a theory is related to the $\alpha =0$ CFT by considering a continuous deformation along the one-parameter family of boundary conditions  $\alpha = \lambda J(x, \beta)$  for $\lambda \in [0,1]$.  The argument below is essentially a Lorentzian version of the argument of \cite{Sever:2002fk}.

Suppose that one deforms some such boundary condition by a small
amount  $\delta \lambda$.  We may compute the corresponding
deformation $\delta S^{FT} = \partial_\lambda S^{FT} \delta
\lambda$ of the dual field theory action using the AdS/CFT version
\cite{Marolf:2004fy}  of the Schwinger variational principle
\cite{Schwing,BryceBook,Bryce} to compute the matrix element of
$\partial_\lambda S^{FT} $ between two states $|\psi_1\rangle,
|\psi_2\rangle$.  Let us define $\hat W_{\alpha,\lambda} (\psi_1,
\psi_2) := \langle \psi_1 | (S_\lambda^{FT} - S_{\alpha
=0}^{FT})|\psi_2 \rangle$.  The Schwinger principle relates the
variation of the inner product $\langle\psi_1 | \psi_2 \rangle$
element to the variation of the action as follows:
\begin{equation}
\label{Schwinger}
\partial_\lambda \hat W_{\alpha,\lambda} (\psi_1,\psi_2) :=  \langle \psi_1 |\partial_\lambda S^{FT}| \psi_2
\rangle = -i
\partial_\lambda\langle \psi_1 | \psi_2 \rangle = \partial_\lambda S^{AdS}_{\psi_1\psi_2},
\end{equation}
where the  function
$S^{AdS}_{\psi_1\psi_2}$ is built from the action $S_{W_\alpha}$ (\ref{SSWa}), together with the bulk wave functions corresponding to the states $|\psi_1\rangle, |\psi_2\rangle.$  Furthermore, the boundary conditions for the variation are such that  $S^{AdS}_{\psi_1\psi_2}$ on the right-hand side of (\ref{Schwinger}) is to be evaluated on the particular solution which causes all $\Sigma^\pm$ boundary terms in $\delta S^{AdS}_{\psi_1\psi_2}$ to vanish \cite{Marolf:2004fy}. This is just the condition that the classical solution considered is the proper stationary point of the path integral to approximate matrix elements between $|\psi_1\rangle$ and $|\psi_2 \rangle$.

As a result, (\ref{Schwinger}) is given just by the terms in $\delta S_{W_\alpha}$ on $\partial {\cal M}$:
\begin{equation}
\label{Schwinger2}
\partial_\lambda \hat W_{\alpha,\lambda} =
   \partial_\lambda B + \int_{\partial \cal M}  \sqrt{\Omega}
(\lambda_+ - \lambda_-) \beta \partial_\lambda \alpha.
\end{equation}
Functionally differentiating this relation with respect to $\beta$ yields:
\begin{equation}
\label{dbeta}
\partial_\lambda  \frac{\delta} {\delta \beta}  \hat W_\alpha =
   \partial_\lambda \frac{\delta B} {\delta \beta} +
\sqrt{\Omega} (\lambda_+ - \lambda_-) \partial_\lambda \alpha +
\int_{\partial \cal M} \sqrt{\Omega} (\lambda_+ - \lambda_-) \beta
\partial_\lambda \frac  {\delta \alpha}{\delta \beta} =
\sqrt{\Omega} (\lambda_+ - \lambda_-) \partial_\lambda \alpha,
\end{equation}
where in the last step we have used (\ref{Beq}) and the rule
$\frac{\delta B}{\delta \beta} = \int_{\partial {\cal M} }
\frac{\delta B}{\delta \alpha(x)} \frac{\delta \alpha(x)}{\delta
\beta}.$

When acting on $\alpha$, the derivative with respect to $\lambda$
produces two types of terms: those associated with the explicit
variation of the form of the boundary condition (\ref{sgen}) which
relates $\alpha$ to $\beta$ as well as an ``implicit'' change
resulting from a possible change in the value of $\beta$ itself.
The point here is that $\beta$ is in general evaluated at some
point between $\Sigma_-$ and $\Sigma_+$, and so must be determined
from the fixed boundary conditions at $\Sigma_\pm$ via the
$\lambda$-dependent dynamics.  As a result, we see that $\hat
W_{\alpha,\lambda}(\psi_1, \psi_2) = W_{\alpha,\lambda} (\beta)$
for a function $W_{\alpha,\lambda}$  whose explicit form satisfies
a version of (\ref{dbeta}) in which the right-hand side is
understood to represent only the explicit change in the form of
$\alpha$. Integrating from $\lambda =0$ to $\lambda=1$, and using
$\alpha_{\lambda=0}=0$ and  $W_{\alpha,\lambda=0}=0$  then yields
\begin{equation}
\label{scalarWa} \frac{1}{\sqrt{\Omega}} \frac{\delta
W_{\alpha,\lambda=1}}{\delta \beta} =  (\lambda_+ - \lambda_-)
\alpha,
\end{equation} so that $W_{\alpha,\lambda=1}$ is just the
potential $W_\alpha$ in (\ref{spot}) which was guaranteed to exist
by (\ref{symcond}). The result (\ref{scalarWa}) gives a version of
the relation from \cite{Witten:2001ua,Berkooz:2002ug} consistent
with the normalizations of (\ref{scalarOa}).

Using large $N$ factorization,  we see from (\ref{scalarOa}) that
\begin{equation}
\label{sSFTa}
\Delta S^{FT} = W_\alpha \big|_{\beta = \frac{1}{\lambda_+-\lambda_-} {\cal O}_\alpha} + {\cal O}(1/N),
\end{equation}
since the matrix elements of the left and right-hand sides agree between any two states $|\psi_1 \rangle, |\psi_2\rangle$, up to $1/N$ corrections.

Similarly, one may show that the field theory action differs from that of the $\beta =0$ CFT by the term
\begin{equation}
\label{sSFTb}
S^{FT} - S^{FT}_{\beta =0}  =  W_\beta \big|_{\alpha = \frac{-1}{\lambda_+-\lambda_-} {\cal O}_\beta} + {\cal O}(1/N),
\end{equation}
where $ W_\beta$ satisfies
\begin{equation}
\label{scalarWb}  \frac{1}{\sqrt{\Omega}} \frac{\delta
W_\beta}{\delta \alpha} = -  (\lambda_+ - \lambda_-) \beta.
\end{equation}

\subsection{Saturating the Breitenlohner-Freedman Bound}
\label{satBF}

Let us now consider the case saturating the Breitenlohner-Freedman
bound, where the asymptotic behavior is
\begin{equation}
\label{BFasympt}
\phi \rightarrow    \frac{\alpha(x) \ln r }{r^{d/2}}  + \frac{ \beta(x)}{r^{d/2}} \, .
\end{equation}

In analogy with (\ref{KWA}), consider the action
\begin{equation}
\label{BFsactiona0} S_{\alpha = 0} = - \int_{\cal M} \left(
\frac{1}{2}\partial \phi^2 + V(\phi) \right) \sqrt{-g} -
\frac{1}{2} \lambda_- \int_{\partial \cal M} \sqrt{-h} \phi^2,
\end{equation}
for which we find
\begin{equation}
\delta S_{\alpha =0} = - \int_{\cal M} \sqrt{\Omega} \alpha( \ln r \delta
\alpha + \delta \beta).
\end{equation}
We see
that $S_{\alpha =0}$ yields a satisfactory variational principle
only for the boundary condition $\alpha=0$.

To fix $\alpha$ to some other value ($\alpha  =  J_\alpha(x)$), we can use
\begin{equation}
S_{\alpha = J_\alpha} = S_{\alpha =0} + \int_{\partial \cal M}
\sqrt{\Omega} \beta J_\alpha.
\end{equation}
Performing the usual calculation then yields
\begin{equation}
\label{scalarOaBF} \langle {\cal O}_\alpha \rangle =
\frac{1}{\sqrt{\Omega}} \frac{\delta S_{\alpha = const}}{\delta
J_\alpha} = \beta.
\end{equation}
Furthermore, if we deform the $\alpha =0$ theory to a theory with
boundary conditions $\alpha = J(x, \beta)$ satisfying
(\ref{symcond}), the arguments of section (\ref{sW}) lead to the
conclusion that the action of the dual field theory has been deformed by
the addition of $W_\alpha({\cal O}_\alpha)$ where
\begin{equation}
\label{scalarWaBF}   \frac{1}{\sqrt{\Omega}} \frac{\delta
W_\alpha}{\delta \beta} =
  \alpha . \end{equation}

In the same way, considering deformations of the $\beta =0$ theory yields
\begin{equation}
\label{scalarObBF} \langle {\cal O}_\beta \rangle =
\frac{1}{\sqrt{\Omega}} \frac{\delta S_{\beta = const}}{\delta
J_\beta} = - \alpha,
\end{equation}
and
\begin{equation}
\label{scalarWbBF}  \frac{1}{\sqrt{\Omega}} \frac{\delta
W_\beta}{\delta \alpha} = -  \beta. \end{equation} However, in
this case the $\beta=0$ theory is not precisely conformal
\cite{Witten:2001ua}. Instead, it has a logarithmic behavior
associated with the $\ln r$ in (\ref{BFasympt}).

\section{Boundary conditions for vector fields}
\label{vecbcsec}

In section \ref{scalar} above, we reviewed the freedom of choosing
boundary conditions for scalar fields.   It is natural to expect
that similar choices of boundary conditions are allowed for
spinors, vectors, and tensor fields in AdS$_{d+1}$ with similar
interpretations in terms of deformations of the dual field theory.
In the scalar case, the range (\ref{mrange}) of masses for which
such boundary conditions are allowed depends on the dimension $d$.
One expects similar results for higher spin fields but, for the
vector and tensor case, we note that one particular value of the
mass (zero, in the obvious convention) will be associated with
gauge invariance. Thus, if one focuses on either vector gauge
fields or the linearized graviton, one expects general boundary
conditions to be allowed only for certain dimensions $d$.  In
fact, such boundary conditions exist for $d=3,4,5$, though only for $d=3$ will they preserve Lorentz invariance.

For simplicity, we focus here on case of a vector field
$A_\mu$ satisfying the source-free Maxwell equation
\begin{equation}
\label{MaxEOM}
\nabla_\nu F^{\mu\nu} = 0,
\end{equation}
though the tensor case is clearly of interest as well. From our
perspective, the fundamental question is what boundary conditions
turn the space of solutions to (\ref{MaxEOM}) into a well-defined
phase space.  Any such setting leads to a well-defined (though not
necessarily renormalizable) framework for perturbative
quantization \cite{WaldThermo,Hollands:2001nf,Hollands:2001fb}. In
particular, we ask under what boundary conditions is the
symplectic structure both finite and conserved, meaning that no
symplectic flux flows outward through the AdS boundary $\partial
{\cal M}$.

\subsection{Symplectic flux through $\partial {\cal M}$}
\label{sympA}

Let us first consider the symplectic flux through a region $R
\subset \partial {\cal M}$ of null infinity. For a Maxwell field,
this is
\begin{equation}
\label{Mflux}
\omega_R(\delta_1 A, \delta_2 A) = - \int_R \sqrt{-h} n^\mu(\delta_1 A^\nu \delta_2 F_{\mu \nu}
-\delta_2 A^\nu \delta_1 F_{\mu \nu}).
\end{equation}
Introducing indices $I, J, K...$ which run over directions in
$\partial {\cal M}$, it is clear that this flux vanishes whenever
the pull-back $A_I$ to ${\partial {\cal M}}$ of $A_\mu$ is
appropriately related to the projection $F^I$ to ${\partial {\cal
M}}$ of
\begin{equation}
F^\nu := - \frac{\sqrt{-h}}{\sqrt{\Omega}} n_\mu F^{\mu \nu}  = -
r^{d} n_\mu F^{\mu \nu},
\end{equation}
where the factor of $-r^{d}$ is chosen to simplify later expressions.  That is, we wish to impose either
\begin{equation}
\label{vecbc}
A_I =  J_{A_I}(x, F\big|_{\partial {\cal M}} ) \ \ \ {\rm or} \ \ \ F^I = J_{F^I} (x, A\big|_{\partial {\cal M}} ),\end{equation}
where
\begin{equation}
\label{symcond}
\frac{\partial J_{A_I}}{\partial F^J} \ \ \ {\rm and} \ \ \  \frac{\partial J_{F^I}}{\partial A_J}
\end{equation}
must be symmetric in order for $\omega_R$ to vanish.  The symmetry
conditions (\ref{symcond}) are just the integrability conditions
for the boundary conditions (\ref{vecbc}) to be specified in terms
of potentials $W_\alpha, W_\beta$ such that
\begin{equation}
\label{AFpot} J_{A_I} =  -\frac{1}{\sqrt{\Omega}}  \frac{\delta
W_A}{\delta F^I}, \ \ \ {\rm or} \ \ \ J_{F^I} =
\frac{1}{\sqrt{\Omega}}  \frac{\delta W_F}{\delta A_I}.
\end{equation}

Since the boundary conditions (\ref{vecbc}) are local on $\partial
{\cal M}$ one expects that these theories are fully local.  In
particular, one expects that the advanced and retarded Green's
functions $G^\pm(x,y)$ vanish unless $x$ and $y$ are connected by
a causal curve.

Before proceeding, let us make a few observations about the
effects of gauge symmetry and charge conservation.  In
(\ref{AFpot}), we considered $W_A$ to be some fixed functional of
an arbitrary vector field $F^I$ on the boundary.  However, due to
charge conservation, $F^I$ is divergence-free on-shell:
\begin{equation}
\label{Fdivfree} {\cal D}_I F^I =0,
\end{equation}
where ${\cal D}^I$ is the covariant derivative on the boundary.
Thus, if one instead considers $W_A$ as a functional of the
on-shell fields, the variations of $F^I$ are constrained by
(\ref{Fdivfree}) and the functional derivatives (\ref{AFpot}) are
ill-defined.  However, the ambiguity is just that associated with
the gauge freedom; under a gauge transformation $A_\mu \rightarrow
A_\mu + \partial_\mu \Lambda$ we have $J_{A, I} \rightarrow A_I +
\partial_I \Lambda$.  Similarly, due to (\ref{Fdivfree}), we must
have ${\cal D}_I J_{F^I}=0$ on shell.  Thus, on shell and when the
boundary condition holds, $W_F$ must be equal (up to boundary
terms at $\Sigma_\pm$) to some gauge-invariant functional of
$A_I$.

\subsection{Normalizability and boundary conditions}
\label{nbc}

We now turn to the question of normalizability of the modes with
respect to the symplectic structure. A related normalizability
criterion was analyzed in \cite{Ishibashi:2004wx} by Ishibashi and
Wald, whose results will be of central use below. The results of
\cite{Ishibashi:2004wx} are stated in terms of a decomposition of
the vector field $A_\mu$ into vector and scalar parts with respect
to some $SO(d)$ symmetry in AdS$_{d+1}$, which we now recall.

\subsubsection{Preliminaries}
We begin by introducing notation in order to recall the results of
\cite{Ishibashi:2004wx} and to reformulate these results in a more
transparent form. One notes that spheres invariant under the
$SO(d)$ symmetry foliate the spacetime, and that the spheres
themselves can be labelled by the coordinates $y^a$, $a=0,1$ with
$y^0=t,y^1=r$.  It is convenient to introduce an associated
two-dimensional metric
\begin{equation}
\widehat{ds}^2 = \hat g_{ab} dy^a dy^b = -(r^2 +1) dt^2 + \frac{dr^2}{r^2+1},
\end{equation}
with metric-compatible covariant derivative $\hat \nabla_a$, and
Levi-Civita tensor $\epsilon_{ab}$ satisfying $\epsilon_{rt}=1$.
On the unit sphere $S^{d-1}$, we introduce coordinates $z^i$, $i=1
\dots d-1$, and we take the metric  and covariant derivative on
the unit sphere to be $\Omega_{ij}$, $D_i$.

It is useful to introduce
orthonormal bases of scalar and vector eigenmodes of the Laplacian
on $S^{d-1}$, satisfying
\begin{equation}
(D^2 + k_S^2) \mathbb{S}_{k_S} = 0,
\end{equation}
\begin{equation}
\label{Snorm}
\int_{S^n} \mathbb{S}_{k_S} \mathbb{S}_{k_S'} = \delta_{k_S, k_S'},
\end{equation}
\begin{equation}
(D^2 + k_V^2) \mathbb{V}_{i,k_V} = 0, \quad \Omega^{ij} D_i \mathbb{V}_{j,k_V}
= 0,
\end{equation}
\begin{equation}
\label{Vnorm} \int_{S^n} \mathbb{V}_{i,k_V}
\mathbb{V}_{j,k_V'}\Omega^{ij} = k_V^2 \delta_{k_V, k_V'},
\end{equation}
where $D^2 = \Omega^{ij} D_i D_j$. The normalization (\ref{Vnorm})
differs from the one used in \cite{Ishibashi:2004wx}, but is
useful to display certain parallels between the vector and scalar
parts.

Using the above bases, one can decompose $A_\mu$ into a vector and
scalar part with respect to $SO(d)$:
\begin{equation}
A_\mu = A_\mu^V + A_\mu^S,
\end{equation}
where
\begin{equation}
\label{Av}
A_\mu^V dx^\mu = \sum_{k_V} \phi_{V,k_V} \mathbb{V}_{i,k_V}dz^i,
\end{equation}
and
\begin{equation}
\label{As}
A_\mu^S dx^\mu = \sum_{k_S} A_{a k_S} \mathbb{S}_{k_S} dy^a + A_{k_S}
D_i \mathbb{S}_{k_S} dz^i.
\end{equation}
Gauge transformations affect only the scalar part; the
gauge-invariant information in the scalar parts is contained in
a scalar mode $\phi_{S, k_S}$ defined by\footnote{Note that such
scalar
  modes are defined only for on-shell field configurations; the form on the right-hand side is
  closed as a consequence of the equation of motion~\eqref{MaxEOM}.}
\begin{equation}
\label{phis} \nabla_a \phi_{S, k_S} = \epsilon_{ab} r^{d-3}
(\nabla^b A_{k_S} - A^b_{k_S}).
\end{equation}
We emphasize here that $\phi_{S,k_S}, \phi_{V,kV}, A_{k_S}$ depend
only on the $y^a$ coordinates; that is, they are fields only on
the two-dimensional quotient space $AdS_{d+1}/SO(d)$.
In~\cite{Ishibashi:2004wx}, it was found that for these two
scalars fall off at infinity as
\begin{equation}
\label{vasympt} \phi_{V, k_V} =  \alpha_{V,k_V } r^{0} +
\beta_{V,k_V } r^{2-d} + {\cal O}(r^{-2}) +  {\cal O}(r^{-d}),
    d\neq 2
\end{equation}
\begin{equation}
\label{sasympt}  \phi_{S, k_S} = \left\{ {\alpha_{S, k_S } r^{d-4}
+ \beta_{S, k_S } r^{0} + {\cal O}(r^{-2}) +  {\cal O}(r^{d-6}) \
\ \ {\rm for} \ \ \ d\neq 4} \atop { \beta_{S, k_S }  + \alpha_{S,
k_S } \ln r + {\cal O}(r^{-2} \ln r)\ \ \ {\rm for} \ \ \ d=4}
\right. \ .
\end{equation}
Note that there are no vector modes for $d=2$, as all vector harmonics with non-zero angular momentum on $S^1$ are the gradients of scalars.

Equations (\ref{vasympt}) and (\ref{sasympt}) are the main results
we take from \cite{Ishibashi:2004wx}, but it will be useful to
summarize these results in a somewhat more local and covariant
form. To this end we construct fields $\alpha_S, \beta_S,
\alpha_i, \beta_i$ on the boundary from the modes
$\alpha_{S,k_S}$, $\beta_{S,k_S}$, $\alpha_{V,k_V}$,
$\beta_{V,k_V}$ as follows:
\begin{eqnarray}
\alpha_S(z^i,t)  : = \sum_{k_S}    \alpha_{S, k_S}      {\mathbb
S}_{k_S}, & \beta_S(z^i,t)  : = \sum_{k_S}    \beta_{S, k_S}
{\mathbb S}_{k_S}, \cr \alpha_i(z^i,t)   := \sum_{k_V} \alpha_{V,
k_V} {\mathbb V}_{i,k_V}, & \beta_i(z^i,t)   := \sum_{k_V}
\beta_{V, k_V} {\mathbb V}_{i,k_V}.
\end{eqnarray}
Similarly, we introduce
\begin{equation}
\phi_S  := \sum_{k_S}    \phi_{S, k_S}  {\mathbb S}_{k_S},
 \ \ \ {\rm and \ the \ ``pure \ gauge" \ field} \ \ \ A (z^i,t,r)  : = \sum_{k_S}    A_{ k_S}(t,r)      {\mathbb
S}_{k_S},
\end{equation}
so that we may write
\begin{equation}
\label{phiSexpand} \phi_S = \left\{ {\alpha_S r^{d-4} + \beta_S
r^0 + {\cal O} (r^{-2}) + {\cal O}(r^{d-6}) \ \ \ {\rm for} \ \ \
d \neq 4} \atop {\alpha_S \ln r + \beta_S r^0 + {\cal O}
(r^{-2}\ln r)  \ \ \ {\rm for} \ \ \ d = 4} \right. \ ,
\end{equation}
\begin{equation}
\label{Ai} A_i =  D_i A + \alpha_i(z^i,t) r^{0} + \beta_i(z^i,t)
r^{2-d} + {\cal O}(r^{-2}),
\end{equation}
\begin{equation}
\label{At} A_t = \partial_t A  + r^{5-d}\hat  \nabla_r \phi_S =
\partial_t A  + c_S(d) \alpha_S + {\cal O}(r^{2-d}) + {\cal O}(r^{-2}),
\end{equation}
and
\begin{equation}
\label{Ar} A_r = \partial_r A  + r^{1-d} \hat \nabla_t \phi_S,
\end{equation}

where
\begin{equation}
\label{cSd}
 c_S(d) = \left\{ {(d-4) \ \ \  {\rm for} \ d \neq 4}
\atop { 1  \ \ \  \ \ \ \ \ \ \ \ \ {\rm for} \ d = 4}
 \right. \ .
 \end{equation}
Furthermore, note that $F_{ab} = \epsilon_{ab} F$ where
\begin{equation}
F = -(1/2) \epsilon^{ab}F_{ab} = - \hat \nabla_a r^{3-d} \hat
\nabla^a \phi_S= D^2 \phi_S r^{1-d}  ,
\end{equation}
and where the last step follows from the equation of motion for
$\phi_S$ (eq. (67) from \cite{Ishibashi:2004wx}). Thus we may
write
\begin{equation}
\label{Ft} F^t = - r^{d} n_\mu F^{\mu t}   = \left\{ { D^2
(\alpha_S r^{d-4} + \beta_S ) + {\cal O}(r^{d-6}) + {\cal
O}(r^{-2})
  \ \ {\rm for} \ \ d \neq 4}
\atop {  - D^2 (\alpha_S \ln r + \beta_S )  +  {\cal O}(r^{-2} \ln
r)
  \ \ {\rm for} \ \ d = 4} \right. \  ,
 \end{equation}
and
\begin{eqnarray} \label{Fi}
 F^{ i} &=& -r^d n_\mu F^{\mu i} = -  \Omega^{ij}  \left[ \hat \nabla_t D_j \phi_S + r^{d-2}  n^\mu \hat \nabla_\mu (A_j - D_j
 A)\right]
 \cr
& =&
 \left\{ { \Omega^{ij}  \left (r^{d-4}  \hat \nabla_t D_j \alpha_S +   \hat \nabla_t D_j \beta_S
 - (2-d) \beta_j \right) + {\cal O}(r^{-2}) + {\cal O}(r^{d-6}) \
 {\rm for} \  d \neq 4  }\atop {
 \Omega^{ij} \left( \hat \nabla_t D_j \alpha_S \ln r+  \hat \nabla_t D_j \beta_S
 - (2-d) \beta_i \right)   +  {\cal O}(r^{-2}\ln r) \
 {\rm for}  \  d = 4
 }
 \right. \  ,
\end{eqnarray}
These results summarize the asymptotic behavior of the gauge field
and form the cornerstone of the normalizability analysis below and
in \cite{Ishibashi:2004wx}.

\subsubsection{Normalizability of the symplectic structure}

The most familiar AdS/CFT boundary conditions for a vector field
are to fix $A_I$ on the boundary~\cite{Witten:1998qj}. From
(\ref{Av}), (\ref{Ai}), (\ref{At}) we see that this corresponds to
fixing $\alpha_{i}$, $\alpha_{S},$ and also the ``pure-gauge"
field $A$.  This is true even for $d=2,3$, where $\beta_{S}$ is
the {\it slower} fall-off part of $\phi_{S k_S}$.  This alone is
enough to make one suspect that more general boundary conditions
should be available, and to motivate a general study.

As stated above, a boundary condition of the form (\ref{AFpot})
will be allowed whenever it renders the symplectic structure
finite. Computing the symplectic structure on a hypersurface
$\Sigma$ defined by $t= constant$ using (\ref{Ai}), (\ref{At}),
(\ref{Ar}), and the fact that the vector modes are divergence-free
on $S^{d-1}$, we find
\begin{eqnarray}
\label{sympnorm}
\omega_\Sigma(\delta_1 A, \delta_2 A) &=& - \int_\Sigma \sqrt{q} \ t^\mu(\delta_1 A^\nu \delta_2 F_{\mu \nu}
-\delta_2 A^\nu \delta_1 F_{\mu \nu}) \cr
&=&-  \int_\Sigma \sqrt{\Omega} d^{d-1}z dr \   r^{d-5} \Omega^{ij} ( \delta_1 \alpha_i + \delta_1 \beta_i r^{2-d}) \hat  \nabla_t ( \delta_2 \alpha_i + \delta_2 \beta_i r^{2-d}) \cr &-& \int_\Sigma \sqrt{\Omega} d^{d-1}z dr \  r^{1-d}  (\hat \nabla_t \delta_1 \phi_S)(D^2 \delta_2 \phi_S) \big] \cr &+&  \int_{\partial \Sigma} \sqrt{\Omega} d^{d-1}z \ \delta_1 A \delta_2 F^t  + (1 \leftrightarrow 2) + {\rm finite},
\end{eqnarray}
where  $t^\mu$ is the unit normal to $\Sigma$ and $q$ is the
determinant of the metric on $\Sigma$. In (\ref{sympnorm}), the
terms implicit in ``finite" come from the higher order corrections
in (\ref{phiSexpand}-\ref{Fi}) and are explicitly finite for $2
\le d \le 6$, which will be the cases of primary interest.

For the vector modes, the inner product
studied in~\cite{Ishibashi:2004wx} agrees with (\ref{sympnorm}) up to a factor of  the mode
frequency $\omega$.  For the scalar modes, the inner product
agrees up to a factor of $\omega$ and a factor of $k_S^2$.
Thus,  the desired normalizability results are directly related to those of ~\cite{Ishibashi:2004wx}:

\begin{itemize}
\item  {\bf d $\le$ 1:} Since the bulk spacetime dimension is $\le
2$, there are no propagating modes for $A_\mu$.  This case is
trivial.

\item {\bf d = 2:} There are no vector modes, and the the
$\beta_{S,k_S}$ modes fail to be normalizable. We therefore choose
to fix $\beta_{S} = J_{\beta_{S}}(x)$  for all $\alpha_{S}$. From
(\ref{Ft}) we see that for $d=2$ the contribution of $\alpha_{S}$
to $F^I$ vanishes at $\partial {\cal M}$.  Thus, fixing
$\beta_{S}$ is equivalent to imposing $F^I\big|_{\partial {\cal
M}} = J_{F^I}(x)$, where $J_{F^I}$ is independent of the dynamical
fields.  We must also keep the pure-gauge field $A$ from growing
too quickly at infinity.  This is easily accomplished by imposing
the gauge condition $\Omega^{ij} {\cal D}_i A_j = {\cal O}(r^2)$.

\item {\bf d = 3:} All modes $\alpha_S, \beta_S,\alpha_i, \beta_i$ are normalizable so long as the pure-gage field $A$ is finite on $\partial {\cal M}$.  Thus, any boundary condition of the form (\ref{vecbc}) is allowed.

\item { \bf d = 4 or 5:}
The $\alpha_{V,k_V}$ modes fail to be normalizable and
must be fixed.  From (\ref{Ai}) we see that, up to gauge
transformations, this is equivalent to imposing
$A_i\big|_{\partial {\cal M}} = J_{A_i}(x)$, where $J_{A_i}$ is
independent of the dynamical fields.

If one considers only the integral over $\Sigma$ in
(\ref{sympnorm}), then all scalar modes are normalizable. However,
because $F^t$ is divergent for $d=4,5$, there is a potential for
the final term involving the pure gauge field $A$ to alter this
conclusion.  We remove this possibility
by noting that the above
boundary condition on $A_i$ fixes $A$ on the boundary and
by also
imposing the gauge condition $\Omega^{ij} {\cal D}_i ( A_j -
J_{A_j}) = {\cal O}(1/r)$. We may then use any boundary condition
of the form
\begin{equation}
\label{45d} A_t =  \frac{1}{\sqrt{\Omega}}  \frac{\delta
W_{A_t}}{\delta F^t} \ \ \ {\rm or} \ \ \ F^t =  -
\frac{1}{\sqrt{\Omega}}  \frac{\delta W_{F^t}}{\delta A_t},
\end{equation}
where $W_{A_t}$ is the integral of a local function of $F^t$ alone
or $W_{F^t}$ is the integral of a local function of $A_t$ alone.

As noted above, $F^t$ is divergent for general values of $\alpha_S$.  Nonetheless, we may display the above boundary conditions in a manifestly finite form by introducing the quantity
$F_{\beta_S=0}^I$, defined by setting $\beta_{S,k_S}=0$ in the
mode expansion (\ref{Ft}), (\ref{Fi}) of $F^I$.  We also introduce
$F_{\beta_S \ only}^I :=  F^I - F_{\beta_S=0}^I$ which is finite
on $\partial{\cal M}$.  We may then reformulate (\ref{45d}) as
\begin{equation}
\label{45d2} A_t =  \frac{1}{\sqrt{\Omega}} \frac{\delta
W_A}{\delta F_{\beta_S \ only}^t} \ \ \ {\rm or} \ \ \ F_{\beta_S
\ only}^t = - \frac{1}{\sqrt{\Omega}} \frac{\delta \tilde
W_F}{\delta A_t},
\end{equation}
where $\tilde W_{F} = W_{F} + F_{\beta_S=0}^tA_t$.  Choosing $W_A$ to be a finite function of
$F_{\beta_S \ only}^t$ or choosing $\tilde W$ to be a finite function of $A_t$ results in a well-defined boundary condition.

\item {\bf d $\ge$ 6:}  Neither the $\alpha_{V,k_V}$ modes nor the
$\alpha_{S,k_S}$ modes are normalizable.  We must impose
$A_I\big|_{\partial {\cal M}} = J_{A_I}(x)$, with $J_{A_I}$
is independent of the dynamical fields.
\end{itemize}

Ishibashi and Wald studied the case of linear boundary conditions
in detail, and obtained interesting results as to which boundary
conditions yield stable bulk theories. In contrast, our desire is
to understand the general boundary condition above in terms of
deformations of the dual field theory.  We turn to this question
in section \ref{cft} below.

\section{Dual CFT description}
\label{cft}

For a scalar field with  $\alpha$ completely fixed by the boundary condition, the expectation value
of the operator dual to deformations of $\alpha$ is given by
$(\lambda_+-\lambda_-) \beta$. The dimension of this operator is thus related to the
scaling of $\beta$ in the bulk spacetime. Similarly, if we fix the value of
$\beta$, the dimension of the operator associated with
variations of $\beta$ is related to the scaling of $\alpha$ in the
bulk spacetime.

Here we study the corresponding relations and the details of the
operators dual to a vector gauge field. At least for $d=3$, we
expect to have two operators ${\cal O}_{A,}{}^I$ and ${\cal O}_{F,
I}$ dual to variations of $A_I$ and $F^I$ respectively.  Now,
under a scaling $r \rightarrow \Lambda r$, the components of the
gauge field scale as $A_I \rightarrow A_I$, while  $F^I
\rightarrow \Lambda^{1-d} F^I$. Thus, ${\rm \ dim} \ {\cal
O}_{A,}{}^I ={\rm \ dim} \  F^I =d-1$, which has  the right
dimension to represent a conserved current.

On the other  hand, ${\rm dim} \ {\cal O}_{F,I} ={\rm dim} \ A_I
=1$.  We note that this agrees with the results of
\cite{Leigh:2003ez} obtained by CFT methods.   At first, this may
seem like a surprisingly low dimension. Indeed, the dimension of
local vector-like observables in a unitary CFT is bounded below by
$d-1$ (see e.g.~\cite{Minwalla:1997ka}). The natural conclusion
\cite{Witten:2003ya,Leigh:2003ez} is that ${\cal O}_{F,I}$ is not
strictly a local {\it observable}, but instead represents a U(1)
vector gauge field in the CFT.

The details of this picture are discussed below.  We present bulk
actions appropriate to each of the boundary conditions stated in
section \ref{vecbcsec} and discuss the corresponding implications
for the dual field theory.    In order to neglect certain
additional terms which contribute in higher dimensions, we
restrict attention to the case $2 \le d \le 5$, which encompasses
the most interesting cases identified above.  The generalization
to higher dimensional cases is straightforward.  We proceed in
parallel with our treatment of the scalar field in section
\ref{scalar}, first reviewing the case where one fixes $A_I$ or
$F^I$ alone, and then considering more general boundary
conditions.

\subsection{Fixing $A_I$ on the boundary}

As noted in section (\ref{nbc}), for $d \ge 3$ we may choose the
familiar boundary condition
\begin{equation}
\label{fixAI} A_I = J_{A_I}(x),
\end{equation}
where $J_{A_I}$ independent of any dynamical fields.  For this
boundary condition, consider the action
\begin{equation}
S_{A=const} = - \frac{1}{4} \int_{\cal M} \sqrt{-g} F_{\mu \nu}
F^{\mu \nu} + \int_{\partial \cal M} \sqrt{-h}    n_\mu
A_\nu F^{\mu \nu}_{\beta_S,\beta_V=0},
\end{equation}
where $F^{\mu \nu}_{\beta_S,\beta_V=0}$ is constructed (in analogy
with $F^{\mu \nu}_{\beta_S=0}$ above) by setting $\beta_{S,k_S} =
\beta_{V,k_V}=0$ in the mode expansion of $F^{\mu \nu}$ for all
$k_S, k_V$.  We also define the analogous
$F^{I}_{\beta_S,\beta_V=0}$.

From (\ref{At}), (\ref{Ai}), (\ref{Ft}), and (\ref{Fi}), it is clear that $F^{I}_{\beta_S,\beta_V=0}$  is a local function (on the boundary) of $A_I \big |_{\partial {\cal M}}$ and its derivatives.
As a result, under a general variation which fixes boundary conditions at $\Sigma_\pm$, we find
\begin{equation}
\label{VarSA} \delta S_{A=const} =  \int_{\partial \cal M} \sqrt{\Omega}
F^I_{\beta \ only} \delta A_I ,
\end{equation}
where $F^I_{\beta \ only} = F^I - F^I_{\beta_S,\beta_V=0}$ and we have used the equations of motion for the background.
Clearly, (\ref{VarSA}) vanishes when the variation preserves (\ref{fixAI}).  The corresponding dual operator ${\cal O}_{A,}^I$ satisfies
\begin{equation}
\langle {\cal O}_{A,}{}^I \rangle = \frac{1}{\sqrt{\Omega}}
\frac{\delta S_{A=const}}{\delta A_I} = F^I_{\beta \  only}.
\end{equation}
Of course, conservation of this current follows from gauge invariance, and it is natural to introduce the notation $j^I = {\cal O}_{A,}{}^I$.  This is the familiar AdS/CFT duality for vector fields \cite{Witten:1998qj}.

\subsection{Fixing $F^I$ on the boundary}

For $d=2$ and $d=3$, we have seen that an allowed boundary condition is to set
\begin{equation}
\label{fixFI} F^I = J_{F^I}(x),
\end{equation}
where $J_{F^I}$ is independent of any dynamical fields.  From
(\ref{Ft}), (\ref{Fi}) we see that, for such values of $d$, the
condition (\ref{fixFI}) fixes $\beta_{S,k_S}$ and $\beta_{V,k_V}$
but leaves $\alpha_{S,k_S}$ and $\alpha_{V,k_V}$ unconstrained.
For $d=2$ this in fact the {\it only} allowed boundary condition in our class.

For the boundary condition (\ref{fixFI}), consider the action
\begin{equation}
S_{F=const} =  - \frac{1}{4} \int_{\cal M} \sqrt{-g} F_{\mu \nu}
F^{\mu \nu} +  \int_{\partial \cal M} \sqrt{-h}    n_\mu A_\nu
F^{\mu \nu} .
\end{equation}

Under a general variation which fixes boundary conditions at $\Sigma_\pm$, we find
\begin{equation}
\label{VarSF} \delta S_{F=const} = -\int_{\partial \cal M}
\sqrt{\Omega}A_I \delta F^I ,
\end{equation}
where we have used the equations of motion for the background. The
result (\ref{VarSF})  vanishes as required when the variation
preserves (\ref{fixAI}).  The corresponding dual operator ${\cal
O}_{F,I}$ satisfies
\begin{equation}
\label{OFI} \langle {\cal O}_{F,I} \rangle =
\frac{1}{\sqrt{\Omega}} \frac{\delta S_{F=const}}{\delta F^I} =
-A_I + \partial_I \Lambda.
\end{equation}
Here $\Lambda$ is an arbitrary function on $\partial {\cal M}$
introduced to take account of the fact that, since (\ref{OFI})
uses the on-shell action, variations of $F^I$ are constrained to
satisfy ${\cal D}_I F^I =0$.  Thus, functional derivatives with
respect to $F^I$ are inherently ambiguous.  This ambiguity
strongly suggests that ${\cal O}_{F,}{}^I$ is itself a vector
gauge field in the dual theory. For $d=3$, this conclusion was
reached previously in \cite{Witten:2003ya,Yee:2004ju} using
related path-integral reasoning.  As observed in
\cite{Leigh:2003ez}, the well-defined (i.e., gauge invariant) part
of ${\cal O}_{F,}{}^I$ is inherently a non-local operator and is
thus not subject to the bound $\Delta \ge d-1$ on the dimension of
local vector operators.

Recall that for $d=2,3$ the engineering dimension of a vector
gauge field is $0,1/2$.  In contrast, ${\rm dim} \ {\cal
O}_{F,I}=1$, so the anomalous dimension of this operator is $1$
for $d=2$ and $1/2$ for $d=3$.  From this point of view, it is no
surprise that there is no $F^I=0$ CFT for $d > 4$; such theories
would necessarily contain operators with negative anomalous
dimension.  The case $d=4$ is clearly marginal, and the $F^I=0$
CFT fails to exist due to the logarithmic behavior at large $r$.

\subsection{More general boundary conditions}

For $d=3$ we may consider any boundary conditions (\ref{vecbc})
determined by some $W_A$ or $W_F$.  A general class of boundary
condition (\ref{45d}) is also available in $d=4,5$.  There we
cannot consider the theory as a deformation of the $F^I=0$ theory
(which does not exist), but it does make sense to define the
theory through any functional $W_{A} = W_{A_t} + \int
\sqrt{\Omega} J_{A_i} F^i$, where $W_{A_t}$ is an integral of a
local function of $F^t$.

Let us therefore consider (in $d=3,4,5$) such a boundary condition
as a deformation of the $A_I = constant$ theory via the action
\begin{equation}
S_{W_A} = S_{A=const}  + B_A(A\big|_{\partial {\cal M}} ).
\end{equation}
It is clear that for this action is to be stationary on solutions we must have
\begin{equation}
\frac{1}{\sqrt \Omega}\frac{\delta B_A}{\delta A_I} = - F^I_{\beta
\ only}.
\end{equation}
It is also clear that $B_A$ is local on the boundary and, since
$F^I$ is conserved, $B_A$ is gauge-invariant at least on-shell.
The same calculation as in section \ref{scalar} now shows that the
deformation of the dual field theory action is the Legendre
transform of $B_A$:
\begin{equation}
\label{SWA} \langle \Delta S^{FT} \rangle  = B_A - \int_{\partial
{\cal M}} \sqrt{\Omega} F^I_{\beta \ only} A_I.
\end{equation}

Assuming that our boundary condition associates every $F^I_{\beta
\ only}$ with some $A_I$, we may regard $\langle \Delta S^{FT}
\rangle$ as a function of $F^I_{\beta \ only}$. One would now like
to functionally differentiate (\ref{SWA}) with respect to
$F^I_{\beta \ only}$. However, since we have worked on-shell, our
expression $\langle \Delta S^{FT} \rangle$ is only defined for
divergence-free vector fields $F^I_{\beta \ only}$. The result is
therefore
\begin{equation}
\frac{1}{\sqrt \Omega}\frac{\delta \langle \Delta S^{FT}
\rangle}{\delta F^I_{\beta \ only}} = - A_I + \partial_I \Lambda.
\end{equation}
Except for the term $\partial_I \Lambda$, this is the equation
(\ref{AFpot}) satisfied by $W_A$. Thus we find $\Delta S^{FT}  =
W_A + constant$ up to a term of the form $\int_{\partial {\cal M}}
\sqrt{\Omega} F^I_{\beta \ only} \partial_I \Lambda$.   Since
$\partial_I F^I_{\beta \ only} =0$ in the large $N$ limit of the
dual field theory, this amounts to the expected statement that
$\Delta S^{FT} = W_A + constant$ up to $1/N$ corrections (and
perhaps a boundary term at $\Sigma_\pm$). The behavior at higher
order in $1/N$ is determined by the structure of gauge anomalies
in the bulk theory.

Similarly, for $d=3$ one may regard a generic boundary condition
as a deformation of the $F^I = constant$ theory via the action
\begin{equation}
S_{W_F} = S_{F=const}  + B_F(F\big|_{\partial {\cal M}} ),
\end{equation}
defined by
\begin{equation}
\frac{1}{\sqrt \Omega}\frac{\delta B_F}{\delta F^I} = A_I +
\partial_I \Lambda,
\end{equation}
where $\Lambda$ is arbitrary.  Since the construction of the dual
field theory deformation proceeds on-shell, this ambiguity in
$B_F$ leads at most to a boundary term at $\Sigma_\pm$.  Again one
finds that the $\langle \Delta S^{FT} \rangle$ is the Legendre
transform of $B_F$.

We wish to regard $\langle \Delta S^{FT} \rangle$ as a functional
of $A_I$.  Because we now work on-shell, simply using the boundary
condition to replace $F^I$ by $A_I$ would define $\langle \Delta
S^{FT} \rangle$ only for those $A_I$ for which the boundary
condition yields divergence-free $F^I$.  Let us therefore consider
only boundary conditions for which {\it every} $A_I$ differs from
some $A_I^{div-free}$ only by a gauge transformation, where
$A_I^{div-free}$ is a connection associated by the boundary
condition to some divergence-free $F^I$.  This is the natural
analogue of the condition imposed above in discussing deformations
of the $A_I = constant$ theory. Since $\Delta  S^{FT}$ must be
gauge-invariant up to boundary terms, our new assumption allows us
to define $\langle \Delta S^{FT} \rangle$ for all $A_I$.  Taking a
functional derivative then shows that for any $A_I^{div-free}$ we
have $\Delta S^{FT} = W_F$, up to an additive constant and the
usual boundary terms at $\Sigma^\pm$. Thus, $ \Delta S^{FT}$ is
just the gauge-invariant version of $W_F$ mentioned at the end of
section \ref{sympA}.

Let us examine the particular case of linear boundary conditions in detail:
\begin{equation}
\label{linear} F^I_{\beta \ only}  = \gamma^{IJ} A_J,
\end{equation}
for some  $\gamma^{IJ}$ with inverse $\gamma_{IJ}$.  (For $d=4,5$
we must have $\gamma_{IJ} \propto \delta_{It} \delta_{Jt}$ and
$\gamma^{IJ}$ does not exist.) Note that all solutions satisfying
(\ref{linear}) will also will satisfy the gauge condition
\begin{equation}
\label{gauge}
\gamma^{IJ} \partial_I A_J =0.
\end{equation}
For $d=3$ we have
\begin{equation}
\label{Why} W_F = \frac{1}{2} \int_{\partial {\cal M}}
\sqrt{\Omega} A_I A_J \gamma^{IJ} = \frac{1}{2} \int_{\partial
{\cal M}} \sqrt{\Omega} A_I ( \gamma^{IJ}  - \Box^{-1}_\gamma
\gamma^{IK}
\partial_K \gamma^{JL} \partial_L) A_J  ,
\end{equation}
where $\Box_\gamma = \gamma^{IJ} \partial_I \partial_J$ and the
inverse is defined using Dirichlet boundary conditions at
$\Sigma_\pm$.  In the last step, we have used the gauge condition
(\ref{gauge}).  Note that this final form of $W_F$ is invariant
under gauge transformations which vanish on $\Sigma_\pm$.

The relevant (${\rm dim} = 2$) operator (\ref{Why}) will generate
a renormalization-group flow away from the $F^I=0$ CFT. The
deformation is non-local when expressed in terms of
gauge-invariant operators, but becomes local in Lorentz gauge.
This is consistent with the fact that the bulk theory in this
gauge satisfies local field equations and a local boundary
condition.  Although there is no $F^I=0$ CFT for $d=4,5$, we will
discuss a similar UV fixed point for $d=5$ renormalization-group
flows (and a logarithmic theory for $d=4$) in section \ref{hybrid}
below.

Of course, we can also describe a general boundary condition as a
deformation of the $A_I=0$ CFT by
\begin{equation}
W_A = \frac{1}{2} \int \sqrt{\Omega} F^I_{\beta \ only} F^J_{\beta
\ only} \gamma_{IJ},
\end{equation}
which is an irrelvant operator of dimension $2d-2$.  As in the
case of scalar fields, it is thus natural to conjecture (for
$d=3$) that the renormalization-group flow from the $F^I=0$ theory
in the UV has an IR fixed point at the $A_I=0$ CFT.  See
\cite{Leigh:2003gk,Leigh:2003ez} for further discussion of such
flows from the CFT point of view.

\subsection{Hybrid Boundary Conditions and their deformations}
\label{hybrid}

As noted above, in $d=4,5$ the boundary conditions $F^I=0$ are not
allowed due to the failure of the vector modes associated with
$\alpha_{V}$ to be normalizable.  However, the scalar modes
$\alpha_{S}$ {\it are} normalizable, and one may consider `hybrid'
boundary conditions of the form
\begin{equation}
\label{hbc} A_i = J_{A_i}(x), \ \ \  F^t_{\beta \ only} =
J_{F^t}(x).
\end{equation}
For $J_{A_i}=0 = J_{F^t}$, these boundary conditions are again
conformal for $d=5$, though for $d=4$ conformal invariance is
broken by the logarithmic dependence on $r$. Furthermore, such
boundary conditions may be deformed to yield any relationship of
the form (\ref{45d2}). These boundary conditions may also be used
in $d=3$, where other hybrid options also exist. For simplicity,
we confine ourselves here to (\ref{hbc}), but the other $d=3$
hybrid boundary conditions can be handled similarly.

Consider the action
\begin{equation}
S_{hybrid} = S_{A=const} - \int_{\partial {\cal M}} \sqrt{-h} A_t
F^t_{\beta \ only}.
\end{equation}
Under a general variation which fixes boundary conditions at
$\Sigma_\pm$, we find from (\ref{VarSA}) that
\begin{equation}
\label{VarSH} \delta S_{hybrid} =   \int_{\partial \cal M}
\sqrt{\Omega} \left(F^i_{\beta \ only} \delta A_i - A_t \delta
F^t_{\beta \ only} \right) ,
\end{equation}
where  we have used the equations of motion for the background.
Clearly, (\ref{VarSH}) vanishes when the variation preserves
(\ref{hbc}). The corresponding dual operators ${\cal O}_{A,}^i$,
${\cal O}_{F,t}$ satisfy
\begin{eqnarray}
\langle {\cal O}_{A,}{}^i \rangle &=& \frac{1}{\sqrt{\Omega}}
\frac{\delta S_{hybrid}}{\delta A_i} = F^i_{\beta \  only}, \cr
\langle {\cal O}_{F,t} \rangle &=& \frac{1}{\sqrt{\Omega}}
\frac{\delta S_{hybrid}}{\delta F^t} = - A_t.
\end{eqnarray}
Here there are no restrictions on $F^t$, so that the functional
derivative $\frac{\delta}{\delta F^t}$ is well-defined.  The
result is a set of local operators.  For $d=5$ these operators
have conformal dimensions ${\rm dim} \ {\cal O}_{A,}{}^i= d-1$ and
${\rm dim} \ {\cal O}_{F,t}=1$.

Much as with the $d=3$ theory with $F^I=0$, for $d=5$ we may
regard the hybrid theory with $J_{A_i} = 0 = J_{F^t}$ as a UV
fixed point which we can deform by relevant operators (such as
$\int_{\partial {\cal M}} \sqrt{\Omega} A_t A_t$) to generate a
renormalization-group flow. Again, we expect that this flow leads
to an IR fixed point corresponding to the $A_I=0$ theory. Although
the hybrid theory breaks Lorentz invariance, we see that Lorentz
invariance is restored at the IR fixed point.

Our hybrid theory also has an interesting class of marginal
deformations. Given any anti-symmetric tensor $\omega_{IJ}$, we
may consider
\begin{equation}
W_{\omega} = \int_{\partial {\cal M}} \sqrt{\Omega} \omega_{it} {\cal O}_{F,t} {\cal O}_{A,}{}^i
=
- \int_{\partial {\cal M}} \sqrt{\Omega} \omega_{it} A_t F^i_{\beta only},
\end{equation}
which leads to boundary conditions related to (\ref{hbc}) by a
Lorentz transformation.  Due to Lorentz symmetry in the bulk, this
operator should be exactly marginal at all orders in $1/N$.

\section{Discussion}
\label{disc}

In this work, we have studied field theories dual to AdS theories
with deformed boundary conditions for vector fields.  Our analysis
used results from~\cite{Ishibashi:2004wx} concerning the
asymptotics of  vector gauge fields in AdS$_{d+1}$ to read off the
general local boundary condition which leads to a well-defined
phase space, and thus to a well-defined quantum theory.  We then
used the bulk action and the Schwinger variational principle to
construct the associated multi-trace deformations of a dual CFT.
The results are qualitatively similar to those obtained for
general scalar field boundary conditions
\cite{Witten:2001ua,Berkooz:2002ug,Sever:2002fk}, which were also
reviewed in detail.

The results are best summarized separately for each dimension $d$.
The cases $d \le 1$ are trivial as vector gauge fields have no
propagating degrees of freedom.

For $d=2$, there is a unique allowed class of local boundary
conditions $F^I = constant$.  In particular, the most familiar
boundary condition $A_I = constant$ is {\it not} allowed, as it would
fix all of the normalizable modes. This can be understood intuitively
by considering the description in terms of a dual potential in the
bulk, as in~\cite{Witten:2003ya}. In three bulk dimensions, this is a
massless scalar field defined by $* F = d \phi$, and $F^I = constant$
corresponds to the usual boundary condition for the scalar, fixing the
slower falloff part.

In the original Maxwell field picture, one expects the dual operator
to be another U(1) vector gauge field, and not the usual R-charge
current.  However, this vector gauge field is a dimension 1 operator
(i.e., its anomalous dimension is 1 as well), and so has the same
dimension as a conserved current.  We also note that the typical
AdS$_{3}$ gauge fields which arise in AdS$_{3}$/CFT$_2$ are not strict
Maxwell fields, but have Chern-Simons terms which in $d=2$ effectively
provide a mixing between $A_I$ and $F^I$.  Clearly, these Chern-Simons
terms should be taken into account in a complete analysis.

The most general boundary conditions arise for $d=3$, and the
results are similar to those for scalar fields near, but slightly
above, the Breitenlohner-Freedman bound. For $d=3$, any local
boundary condition relating $A_I$ and $F^I$ is allowed, so long as
it is determined by a potential, see (\ref{AFpot}).  We find
Lorentz invariant CFTs associated with the boundary conditions
$A_I=0$ and $F^I=0$, and any linear boundary condition is
associated with a renormalization-group flow from the $F^I=0$
theory (the UV fixed point) to the $A_I=0$ theory (the IR fixed
point).

As in the case of $d=2$, the dual operator in the $F^I=0$ theory
is a vector gauge field with conformal dimension $1$. Using the
associated gauge freedom, the relevant operators that generate
such renormalization-group flows can be expressed in two distinct
ways. When expressed in a gauge-invariant form, the operator is
non-local.  However, with the gauge condition implied by the
general boundary condition, the operator is completely local. This
is consistent with the fact that the bulk theory in this gauge
satisfies local field equations and a local boundary condition. In
particular, the bulk advanced and retarded Green's functions
$G^\pm(x,y)$ vanish unless $x$ and $y$ are connected by a causal
curve.  Since the supports of advanced and retarded Green's
functions in the CFT are given by the boundary limits of those for
the bulk Green's function, we see that the CFT satisfies the usual
notion of causality in this gauge.

In the case $d=4,5$, one must fix the vector part of $A_I$, and there
is no $F^I=0$ theory.  However, the scalar part still admits a variety
of boundary conditions.  For $d=5$, this leads to a new `hybrid' CFT
defined by the boundary conditions $F^t=0, A_i=0$, which explicitly
break Lorentz invariance.  This CFT is a UV fixed point for
renormalization-group flows that lead to the $A_I=0$ CFT where Lorentz
invariance is restored\footnote{This hybrid CFT and others like it
  also exist for the case $d=3$.}. For $d=4$ such boundary conditions
lead to a logarithmic field theory. For $d \ge 6$, only the $A_I=0$
theory is allowed.

Since we consider only gauge fields (which necessarily have
vanishing mass),  the dimension dependence above reflects the fact
that, in the case of scalar fields, the freedom to choose
non-trivial boundary conditions depends on the relation between
the mass $m$ and the dimension $d$.  In that case one understands
the allowed range (\ref{mrange}) in terms of the unitarity bound
$\Delta \ge (d-2)/2$ on the conformal dimension of scalar
operators.  If a CFT with `conjugate' boundary conditions were
allowed for scalars with mass above the upper boundary of
(\ref{mrange}), it would contain an operator violating this bound.
Hence, it does not exist\footnote{The case where the upper bound
of (\ref{mrange}) is saturated and $\Delta = (d-2)/2$ is clearly
marginal.  In principle such a CFT is allowed, but the
corresponding anomalous dimension would have to vanish.  Since for
this case normalizability fails in the bulk, one expect that there
is no such AdS/CFT correspondence.}.  We see that the picture here
is similar:  any $F^I=0$ CFT would contain a vector gauge field of
conformal dimension $1$.  If such a theory were to exist for $d >
4$, the corresponding operator would have negative anomalous
dimension.  The case $d=4$ is a marginal special case.  It would
be interesting to determine if the failure of the $A_I=0$ theory
for $d=2$ and the failure of the hybrid theories for $d > 5$ can
be understood in a similar way.

In the above, we considered a free Maxwell gauge field.  It is
interesting, however, to extrapolate our results to more
complicated cases.  For simplicity, we focus on the case $d=3$.
One immediate generalization is to the $SO(8)$ non-abelian gauge
fields of AdS$_{4}$ supergravity \cite{deWit:1981eq,deWit:1982ig}.
As mentioned in \cite{Witten:2003ya}, 
one expects that the asymptotics and thus the boundary conditions
are governed by the linear theory, and that there is again a UV
CFT dual to the boundary conditions $F^{IA}=0$, where $A$ is an
adjoint $SO(8)$ index.  This CFT appears to contain an $SO(8)$
gauge field in addition to the usual $SU(N)$ gauge field. In some
sense, the usual R-symmetry has been gauged. 

Our results for
vector gauge fields were based heavily on the analysis of
Ishibashi and Wald~\cite{Ishibashi:2004wx}, who also analyzed
boundary conditions for rank 2 tensor fields in the bulk; i.e.,
for the linearized graviton.  Again for this case, very general
boundary conditions were allowed for $d=3$.  Extrapolating our
results above, we therefore predict a new Lorentz-invariant
AdS$_4$/CFT$_3$ correspondence where the graviton satisfies
`conjugate' boundary conditions in the bulk.  With the usual
boundary conditions, the graviton is dual to the CFT stress-energy
tensor.  However, for the conjugate boundary conditions the bulk
graviton must be dual to a spin-2 operator with spin-2 gauge
invariance; i.e., the CFT$_{3}$ is in fact a quantum gravity
theory!  A similar observation was made in
\cite{Leigh:2003gk,Leigh:2003ez} working from the CFT side.  It is
reassuring that quantum gravity in $d=3$ is a finite theory
\cite{PZ,Witten:1988hc,Deser:1989xu,Ooguri:1991ib} due to the lack
of propagating degrees of freedom for the graviton \cite{DJtH,DJ}.
For $d=4,5$ we expect hybrid theories of what might still be
called `quantum gravity,' but which break (local) Lorentz
invariance.

A further generalization would be the inclusion of supersymmetry.
The theories discussed above, and those dual to deformations of
bulk scalars, are not super\-symmetric because they include no
corresponding deformations of the Fermions. However, one expects
the allowed boundary conditions for bulk spinor fields to be
qualitatively similar to those for fields of integer
spin,
with appropriate combinations providing super-symmetric theories.
We therefore conjecture that the `conjugate' AdS$_4$/CFT$_3$
duality described above (with quantum gravity in the CFT) can be
taken to be maximally supersymmetric.

Finally, one may ask about the stability of such exotic theories.
Since such stability should be guaranteed by supersymmetry,
stability itself may be taken as a test of the self-consistency of
the above conjectures.  At the linearized level for fields of spin
0,1,2, this question was fully analyzed for the dynamical modes by
Ishibashi and Wald~\cite{Ishibashi:2004wx}.  Interpreting their
results in our language, the $F^I=0$ and hybrid theories are
indeed linearly stable.

After posting the first version of this paper on the hep-th arxiv, we
became aware of a body of literature containing the main results
presented here for the case $d=3$. In particular, the fact that
`conjugate' boundary conditions in AdS$_4$ are dual to a CFT$_3$ with
a dynamical gauge theory was described in \cite{Witten:2003ya} and
discussed further in
\cite{Leigh:2003gk,Leigh:2003ez,Petkou:2004nu,Yee:2004ju}. Our work
extends this earlier work to other dimensions, and introduces the notion of hybrid boundary conditions.

The perspective we take is also rather different from that of this
previous work. Whereas \cite{Witten:2003ya} started from the CFT
description, and focused on the action of $SL(2,Z)$, we have started
from the bulk spacetime description, and considered all the possible
boundary conditions such that the symplectic flux (\ref{sflux})
vanishes through any region $R$ on the boundary. That is, we start
from a fixed notion of the bulk gauge potential and a fixed form of the symplectic structure, 
and then consider all the
allowed boundary conditions for this formulation of the bulk theory.  In contrast, 
the approach of \cite{Witten:2003ya}  was to consider the usual $A_I =0$ boundary condition for the
different notions of the bulk gauge potential related by
$SL(2,Z)$ and thus to derive boundary conditions on the original gauge potential. Note that the symplectic flux defined by the analogue of
\eqref{sflux} for an $SL(2,Z)$-transformed gauge potential $\tilde
A_\mu$ will in general differ by a boundary term from the one we used
here.    As a
result, some of the boundary conditions $\tilde A_I =0$ will not
preserve our choice of symplectic flux. 
However, the boundary term in the symplectic structure is just that associated with the addition to the action of a Chern-Simons boundary
term, constructed in general from both the vector potential and the dual magnetic vector potential.
Thus, so long as one is
careful to include boundary terms in the action which provide an
appropriate definition of symplectic flux, one can impose
a general boundary condition in terms of any formulation of the bulk
theory: for example, the general boundary condition $\epsilon^{IJK}
{\cal D}_I A_J = \lambda^K_J F^J$ imposed by
\cite{Witten:2003ya,Leigh:2003gk,Leigh:2003ez,Petkou:2004nu,Yee:2004ju}. As one would expect, differences in perspective do not change the physics.

\textbf{Acknowledgements:} We would like to thank David
Berenstein, Stefan Hollands, Sean Hartnoll, Marc Henneaux, Rob
Leigh, Andreas Karch, Per Kraus, Peter Ouyang, Anastasios  Petkou,
Joe Polchinski, Matt Strassler, and Ho-Ung Yee for interesting
discussions related to this work. We would especially like to
thank Akihiro Ishibashi for discussing with us the details of his
analysis of linear boundary conditions in AdS$_{d+1}$. This
research was supported in part by the National Science Foundation
under Grants No. PHY99-07949, No. PHY03-54978, by funds from the
University of California, and by the EPSRC.

\bibliographystyle{utphys}

\bibliography{vectorbc10}

\end{document}